\documentclass[sigconf]{acmart}

\definecolor{myblue}{HTML}{2e21c6}

\usepackage{amsmath,amsfonts}
\usepackage{algorithmic}
\usepackage{graphicx}
\usepackage{textcomp}
\usepackage{xcolor}
\usepackage{siunitx}
\usepackage{subcaption}
\usepackage{makecell,multirow}

\renewcommand\footnotetextcopyrightpermission[1]{} 

\AtBeginDocument{%
  \providecommand\BibTeX{{%
    \normalfont B\kern-0.5em{\scshape i\kern-0.25em b}\kern-0.8em\TeX}}}

\setcopyright{None}
\copyrightyear{0}
\acmYear{0}
\acmDOI{}

\acmConference[]{}{}
\acmSubmissionID{}

\begin{document}


\title{Vector Autoregression in Cryptocurrency Markets: Unraveling Complex Causal Networks}

\author{Cameron Cornell}
\email{cameron.cornell@adelaide.edu.au}
\orcid{1234-5678-9012}
\affiliation{%
  \institution{The University of Adelaide}
  \streetaddress{P.O. Box 1212}
  \city{Adelaide}
  \state{South Australia}
  \country{Australia}
  \postcode{5000}
}

\author{Lewis Mitchell }
\email{lewis.mitchell@adelaide.edu.au}
\orcid{1234-5678-9012}
\affiliation{%
  \institution{The University of Adelaide}
  \streetaddress{P.O. Box 1212}
  \city{Adelaide}
  \state{South Australia}
  \country{Australia}
  \postcode{5000}
}

\author{Matthew Roughan}
\email{matthew.roughan@adelaide.edu.au}
\orcid{1234-5678-9012}
\affiliation{%
  \institution{The University of Adelaide}
  \streetaddress{P.O. Box 1212}
  \city{Adelaide}
  \state{South Australia}
  \country{Australia}
  \postcode{5000}
}

\renewcommand{\shortauthors}{Cornell, Mitchell and Roughan}

\begin{abstract}
Methodologies to infer financial networks from the price series of speculative assets vary, however, they generally involve bivariate or multivariate predictive modelling to reveal causal and correlational structures within the time series data. 
The required model complexity intimately relates to the underlying market efficiency, where one expects a highly developed and efficient market to display very few simple relationships in price data. This has spurred research into the applications of complex nonlinear models for developed markets. 
However, it remains unclear if simple models can provide meaningful and insightful descriptions of the dependency and interconnectedness of the rapidly developed cryptocurrency market. 
Here we show that multivariate linear models can create informative cryptocurrency networks that reflect economic intuition, and demonstrate the importance of high-influence nodes. 
The resulting network confirms that node degree, a measure of influence, is significantly correlated to the market capitalisation of each coin ($\rho=0.193$). 
However, there remains a proportion of nodes whose influence extends beyond what their market capitalisation would imply. 
We demonstrate that simple linear model structure reveals an inherent complexity associated with the interconnected nature of the data, supporting the use of multivariate modelling to prevent surrogate effects and achieve accurate causal representation. 
In a reductive experiment we show that most of the network structure is contained within a small portion of the network, consistent with the Pareto principle, whereby a fraction of the inputs generates a large proportion of the effects. 
Our results demonstrate that simple multivariate models provide nontrivial information about  cryptocurrency market dynamics, and that these dynamics largely depend upon a few key high-influence coins. 

\end{abstract}



\maketitle

\newpage
\section{Introduction}
The emergence of cryptocurrencies has contributed to the expansion of financial opportunities, adding a new and rapidly growing asset class to the economic landscape \cite{joitmc6040197}. These digital assets have attracted 
attention due to their immense market capitalization, high volatility, and decentralized nature. The prices of cryptocurrencies are subject to a complex interplay of factors, such as market sentiment, news events, technological advancements, regulatory changes, and the behavior of market participants. 
The task of modeling cryptocurrency prices is further challenged by the ever-changing market landscape and the complex interconnections between various cryptocurrencies. In response to these challenges, the analysis and modeling of cryptocurrency prices has emerged as vital areas of research for academics, policymakers, and investors alike \cite{regchallenge}. In particular, understanding the interdependencies within the cryptocurrency market has become a focal point in this research arena, serving as a crucial step for accurate representation and intuition of market dynamics.

Financial data modeling encompasses an array of approaches tailored to specific applications and objectives. Many academics and investors study market data, aiming to generate out-of-sample forecasts that may result in profitable trading strategies. Others study simultaneous relationships between asset prices for risk management and portfolio construction. In this study, we concentrate on forecasting methodologies with the aim of inferring networks of inter-dependency and causality in cryptocurrency price data. Our primary objective is to determine whether the resulting network can enhance our understanding of market dynamics and pinpoint subnetworks of high influence. 

A common and widely-accepted approach for multivariate forecasting is the Vector Autoregression (VAR) framework \cite{VAR1, VAR3}. This methodology extends the univariate autoregression technique to account for the cross-dependencies among multiple time series, making it a suitable tool for capturing the complex dynamics between cryptocurrencies. In this paper, we apply the VAR framework to model the price network of 261 different cryptocurrencies, utilising hourly prices over the study period of 1/1/2020 to 1/1/2021, sourced from CoinMarketCap.com.  

Alongside the VAR model, we present two alternate linear causal network constructions for comparison: a correlational network, which does not condition on other system variables, and a Transfer Entropy (TE) network, which conditions on autoregressive effects. Comparatively, our VAR model offers a comprehensive cross-conditioning of system variables.
\newline
\newline
\newline
\newline
The primary contributions of this paper are:
\begin{itemize}
  \item A demonstration that correctly applied linear models can create informative causal networks in cryptocurrency market data.
  \item A comprehensive comparison of the different approaches to forming linear causality networks, including both analytical discourse and a comparative study of the empirical outcomes.  
  \item An investigation of the concentration of information within speculative asset markets, comparing whether the concentration of influence reflects levels of wealth concentration, measured in terms of market capitalization.Quantified by Gini co-efficient, our results indicate that influence (Gini = 0.48) does not exhibit concentration to the extremes shown in the capitalisation distribution (Gini = 0.96). Nevertheless, there does exist a concentrated subset, with 20\% of nodes containing $\sim56\%$  of the total influence, and the majority of significant correlations.
  
\end{itemize}

\begin{figure}[h!]
     \centering
     \includegraphics[width=0.8\columnwidth]{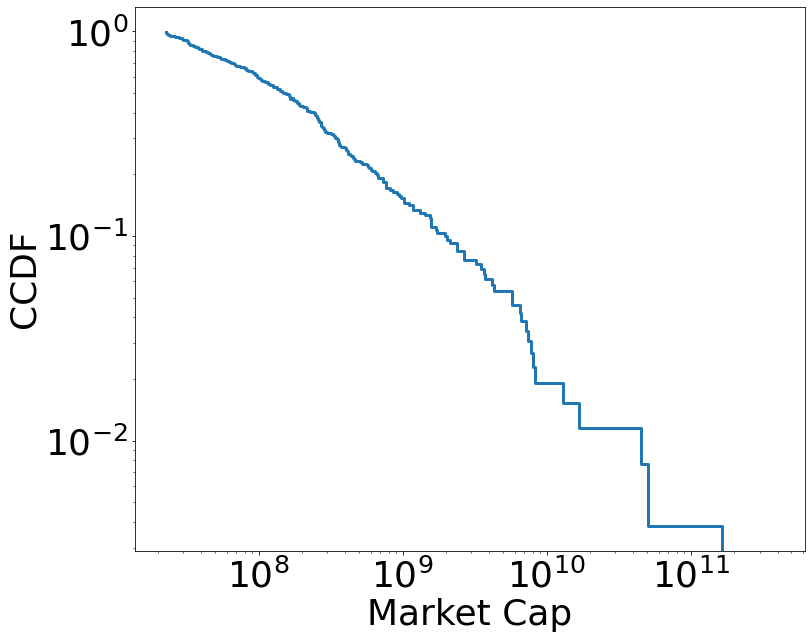}
     \vspace{-0.25cm}
     \caption{Log-log Market Capitalisation CCDF}
     \label{fig:market_cap_ccdf}
     \end{figure}

\begin{figure}[h!]
    \centering
    \includegraphics[width=\columnwidth]{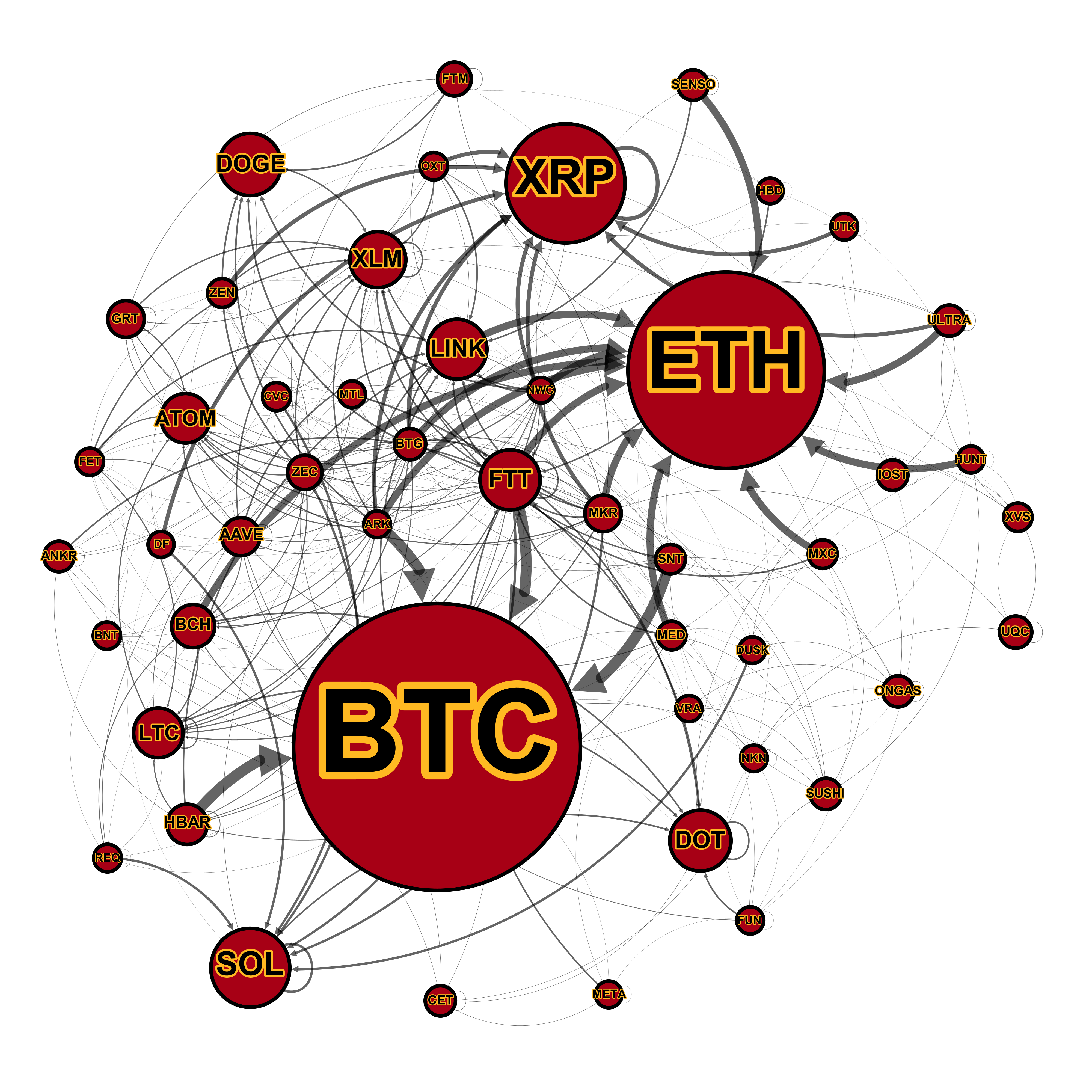}
    \vspace{-1cm}
    \caption{The high influence VAR Subnetwork. Comprising the top 20\% most influential nodes in the original network.}
    \label{fig:network}
\end{figure}

\section{Related work}
Networks based on correlational structures found within time series data of financial assets have been explored in many papers. Most frequently, authors investigate simultaneous correlations between prices, seeking to model the joint structure of these observations as a network  \cite{stat_analysis_of_network}. Several studies track the evolution of these networks over time using sliding window approaches \cite{partial_corr}\cite{structural_entropy}.
The evolution of financial networks has also been investigated by sequentially introducing nodes based on their correlations, and assessing the development of an ``asset graph'' \cite{clustering_and_info}.  

In this paper we investigate cross-correlational effects, which generate causality networks. These types of networks have been constructed before \cite{billio2011}, frequently used to model the joint causal dynamics of price and sentiment \cite{connecting_emotions} \cite{Combiningsocial_and_financial}.

Regardless of whether the context is simultaneous or cross correlational analysis, a common extension is to consider partial-correlations, rather than bi-variate analysis. This extends the graphical representation to show dependency between assets while controlling for the effects of other assets within the dataset. This has be done simply in terms of the partial correlation matrix for simultaneous price modelling \cite{partial_corr}, however for causal modelling this is generally accomplished by variants of the VAR model, which is proportional to partial correlations of the cross-correlation matrix. Variations of VAR models have been developed to incorporate specific features, such as long range dependency 
\cite{long_range_VAR} and restricting the graphical form to acyclic graphs \cite{ahelegebey}. The incorporation of sentiment data has also been investigated for VAR networks \cite{Ahelegbey2021NetworkBE}. While partial correlation and VAR networks condition on all other price variables, other forms of conditioning have been explored, such as conditioning on a specific set of variables meant to represent general market conditions \cite{Nicola2020InformationNM}. 

These studies provide numerous demonstrations of application of VAR methodology to traditional financial data, however there is comparatively almost no literature on the application of these methods to cryptocurrency data. Cryptocurrency causal network studies have generally focused on the interactions between cryptocurrencies and other data forms (sentiment, traditional financial assets etc) \cite{connecting_emotions,crypto1, amirzadeh_nazari_thiruvady_ee, Elsayed2020CausalityAD, AZQUETAGAVALDON2020122574} and frequently focus on bi-variate types of analysis. 
This study demonstrates the application of VAR to the cryptocurrency domain, highlighting it's advantages in controlling for spurious regression effects. A significant focus of this study is to investigate whether multivariate frameworks (VAR/partial correlation analysis) produce tangibly differing results from bi-variate analysis, and to elucidate the mechanics that give rise to this dissimilarity. 

\section{Data}
\label{sec:Data}
The dataset for this study is a collection of hourly cryptocurrency returns from 261 different cryptocurrencies over the time period of January 1, 2021, to January 1, 2022. The selection process started with an attempt to download data on the 750 coins with the highest capitalization. However, a significant number of these coins, mainly those with lower capitalization, had incomplete price histories on an hourly basis. Excluding such cases resulted in slightly above one third of the coins forming viable data.Quantified in terms of total cryptocurrency market capitalisation our data set contains around $79.6\%$ (836B of 1.05T) of the total market.

The returns $x_t$ are generated by taking the logged ratio of subsequent observations in the original price series $p_t$, quoted in terms of the Coin/USD relation. Specifically, the hourly returns are calculated as $x_t = \log\left(p_t/p_{t-1}\right)$. 

In addition to return data, we have gathered the Market Capitalisation (June 2022 figures) for each one of our cryptocurrencies, which indicates the cumulative valuation of the given asset (price per coin $\times$ number of coins). Figure \ref{fig:market_cap_ccdf} displays the complementary cumulative distribution function ( CCDF ) for the capitalisation of  our coins.Interpreting these plots requires understanding fat-tailed distributions, which are typically defined as densities exhibiting significant Skewness or Kurtosis relative to a normal distribution. The most extreme case of this is generally identified as a power-law distribution, where the survival function approaches a power law: $P(X>x)\sim x^{-\alpha}$ as $x \xrightarrow{} \infty$ for $\alpha>0$. A log-log scaled CCDF helps us evaluate such relationships, as it would be visible as a straight line: $\log[P(X>x)] \sim -\alpha \log[x]$. Returning to Figure \ref{fig:market_cap_ccdf}, we observe very slow decay, with approximately linear form. A question we may ask after seeing the significant power-law effect within the CCDF of market capitalisation is 
whether the distribution of influence in our network displays similar scaling, which would be consistent with a scale free network model \cite{scale}.

Table \ref{table:node_attributes} displays several properties of our return series and their correlation to the market capitalisation. The attributes for each coin $i$ include the mean  $\mu_i$ and standard deviation $\sigma_i$ of returns over the study period, as well as the Kurtosis of the return distribution and the Shapiro-Wilk score, which gauges normality \cite{Shapiro1965AnAO}. Additionally, we include the market capitalization, quantified in terms of billions of USD. For correlational analysis of our node attributes we use the Spearman rank correlation, $\rho$ , which measures general monotonic relationships while being robust to outliers. This is crucial for our analysis as the distributions of these attributes are often fat tailed (like capitalisation). The Spearman rank correlation also provides a p-value for testing the null hypothesis of series independence.

The observed Spearman values in Table \ref{table:node_attributes} indicate that lower market capitalisation is correlated with increased variance, elevated Kurtosis and reduced Shapiro-Wilk normality score \cite{Shapiro1965AnAO}. These findings imply high capitalization coins demonstrate greater distribution stability, and low capitalized coins are prone to higher variance and more non-normal, extreme events.

\begin{table}[h!]
    \centering
    \caption{Attributes of Cryptocurrency Return Series: Displaying mean, median, standard deviation $\sigma$, Spearman's rank correlation ($\rho$) with market capitalisation, and associated \textbf{p} value for capitalisation independence.}
    \vspace{-0.25cm}
    \label{table:node_attributes}
    \def\arraystretch{1.1}
    \resizebox{\columnwidth}{!}{
    \begin{tabular}{l S[table-format=3.4] S[table-format=1.4] S[table-format=3.4] S[table-format=1.3] S[table-format=1.4]} 
    \toprule
    \bf{Attribute} & \bf{mean} & \bf{median}  & $\sigma_{attribute}$ & $\rho$ & \bf{p}\\
    \midrule
    $\mu_i$ & 0.0153  & 0.0136 & 0.0181 & 0.203 & \num{9.23e-4} \\ 
    $\sigma_i$  & 3.24 & 2.18 & 4.16 & -0.453 & \num{1.25e-14} \\ 
    Kurtosis & 159 & 31.9 & 656 &  -0.136 & 0.0274   \\
    Shapiro-Wilk & 0.754 & 0.826 & 0.196 & 0.291 & \num{1.78e-6} \\ 
    Capitalisation  & 3.19 & 0.156 & 25.9 &  &  \\
    \bottomrule
\end{tabular}
}
\vspace{-0.5cm}
\end{table}
\section{Constructing networks}
\label{sec:Methodology}
For a given dataset the main methodological choice in developing financial networks is the selection of a suitable model to represent and test the relationships among the price series of different assets (nodes in our graph). The purpose of this section is to provide an overview of our approach to analyzing these interdependencies and to explore two commonly employed alternative methods. Additionally, we will demonstrate how these statistical findings are utilized to generate the network structure.

We aim to construct a directed graph (digraph) denoted by $G=(V,E)$, where V is a set of nodes (Assets) and E is a set of edges (directed causal dependencies). Each edge $e_{ij}$ indicates that the next observation of $j$ depends on the previous values of $i$. The set of edges E may be represented as an $N$x$N$ adjacency matrix W ($N$=$|V|$, the number of assets), with elements $W_{i,j}=1$ if there is a link between $i$ and $j$, with $W_{i,j}=0$ otherwise. 

\subsection{Vector autoregression}
Vector autoregression (VAR) is a popular statistical model introduced by the macroeconometrician Christopher Sims \cite{sims_1980} to model the joint dynamics and causal relations among a collection of time series. It is the natural multivariate extension of the univariate autoregression (AR) model frequently used to analyse the inter-temporal dependency of a sequence of observations. Under the VAR(p) formulation the expectation of the data vector at the next observation is a linear function of $p$ previous observations. Equations 1 and 2 below show the relationship for order 1 and p lagged variants within a system of $N$ variables:  
\begin{align}
\boldsymbol{y}_t &=A_1\boldsymbol{y}_{t-1}+\boldsymbol{c}+\boldsymbol{\epsilon}_t , \\
\boldsymbol{y}_t &=A_1\boldsymbol{y}_{t-1}+A_2\boldsymbol{y}_{t-2}+...+A_{t-p}\boldsymbol{y}_{t-p}+\boldsymbol{c}+\boldsymbol{ \epsilon }_t , 
\end{align}

\noindent where $y_t$ is a $N \times 1$ vector of observations at time $t$, $c$ is a constant term, the $A_k$ are $N \times N$ coefficient matrices for lags $k=1,...,p$, and $\epsilon_t$ is a $N \times 1$ vector of error terms with zero mean and some covariance matrix $\Sigma_\epsilon$, which is often restricted to a Gaussian distribution. The VAR model assumes that the current value of each variable depends on its past values as well as the past values of all other variables in the system (full conditioning). 


The estimation of a VAR model comprises estimating the coefficient matrices $A_k$ and the error covariance matrix $\Sigma_\epsilon$. As we are interested in the causal influence structure within our dataset, we primarily require estimates of $A_k$, as they fully characterise the causal relations. This is often accomplished by the multivariate least squares (MLS) approach under which estimating the VAR is viewed as a general multivariate regression problem, with closed-form solutions generated via orthogonal projection \cite{luet}.

We may conduct hypothesis tests for the statistical significance of the elements of the coefficient matrices by noting that our estimates $\hat{A}_k$ are asymptoptically normally distributed under finite variance assumptions, i.e. ,  
\begin{align}
    \sqrt{N}Vec(\hat{A}_k-A_k)\xrightarrow[]{d}\mathcal{N}(0, \Gamma^{-1}\otimes\Sigma_\epsilon) , \label{eq:kron}
\end{align}
where $\Gamma=\ YY'/N$, $\otimes$ indicates the Kronecker product and $Vec()$ denotes the Vec operation, casting matrices into vector form. For the case of a VAR(1) model the term Y indicates the matrix representation of our response data, implying that $\Gamma$ is the covariance matrix of returns. For generalised VAR(p) the complexity of $Y$ increases, however the result of Equation \ref{eq:kron} remains correct. Hence, we may construct $t_{i,j}$ values associated with the null hypothesis $A_{k,i,j}=0$ as $t_{i,j}=\hat{A}_{k,i,j}/\hat{s}_{i,j}$, with $\hat{s}_{i,j}$ being the relevant term from $\Gamma^{-1}\otimes\Sigma_\epsilon$. To simplify the estimation process and facilitate an analytical comparison with alternative network structures we restrict our investigation to VAR(1) processes. A limitation of this assumption is that it may fail to detect a lag $k>1$ causal effect $i\rightarrow j$ in the absence of a corresponding lag 1 causal effect $i\rightarrow j$. Nevertheless, it is intuitive to assume that such situations would be relatively rare. In light of these considerations, the remainder of this paper will focus on a lag 1 analysis, and we will accordingly omit the index $k$ in our discussions ($A=A_1$).

\subsection{Cross correlation}

A simple approach to assessing the interdependency in lagged price series is to 
consider the bivariate Pearson correlation of the coin returns against the lagged returns of other coins. Under this network construction an edge $e_{i,j}$ is present when a statistically significant $Corr(Y_{i,t-1}, Y_{j,t})=R(1)_{i,j}$ is present. To simplify this procedure, we may note that correlation is intimately linked to covariance, with correlation being a normalised representation of the covariance values. In fact, hypothesis testing the statistical significance of Pearson correlations is equivalent to testing covariances. Hence, we can apply our understanding of the cross covariance function to analyse the expected results of our correlational network if we assume the data was following a VAR process. 

Under the VAR model the autocovariance function of lag 1, commonly denoted $\Gamma(1)$ has an explicit relation to the non-lagged covariance function,  $\Gamma(1)$ = $A\Gamma(0)$. By expanding $\Gamma(0)$ into the following form 
\begin{align}
    \Gamma(0) &=A\Gamma(0)A^T+\Sigma_{\epsilon}. \label{eq:1}
\end{align}
We see that $\Gamma(0)$ has the form of a discrete Lyapunov equation, which under stability conditions has the the following solution:
\begin{align}
    \Gamma(0) &= \sum_{i=0}^{\infty}{A^{i} \Sigma_{\epsilon} {({A}^{T}})^{i} } .   \label{eq:sol}
\end{align}
Combining equation \ref{eq:sol} with the identity $\Gamma(1)$ = $A\Gamma(0)$ allows us to see the lag 1 autocovariance as a function of $A$ and $\Sigma_{\epsilon}$.
\begin{align}
    \Gamma(1) &= \sum_{i=0}^{\infty}{A^{i+1} \Sigma_{\epsilon} {{(A}^{T})}^{i} ,  }
\end{align}

\noindent While this form does not provide immediate insight into the expected structure of $\Gamma(1)$, we can make the key insight that this matrix is a function of both the causal structure $A$, as well as the residual (simultaneous) covariance structure $\Sigma_{\epsilon}$. Hence, it will be an amalgamation of causal and simultaneous dependency. The distinction between these two aspects is critical for financial application, as we often see strong simultaneous dependence and weak causal relations. Hence, it's possible the contributions from $\Sigma_{\epsilon}$ may heavily obscure the perceived causal relations when viewing $\Gamma(1)$. 

To asses the statistical significance of our observed lag-correlations we construct $t_{i,j}$ values associated with the hypothesis $R(1)_{i,j} = 0$.
\noindent For large samples it can be shown that the quantity displayed in Equation \ref{eq:corr} has an asymptotically normal distribution under the assumption of zero correlation \cite{kendall_stuart_1973}. Hence,
\begin{align}
    t_{i,j}=R(1)_{i,j}\times \sqrt{\frac{N-2}{1-R(1)_{i,j}^2}}\xrightarrow[]{d}\mathcal{N}(0, 1) .  \label{eq:corr}
\end{align}

\subsection{Transfer entropy networks}
Another common approach to assess the causal relationships between time series is to measure the information flow between the series, generally assessed in the form of transfer entropy (TE), which is a measure of conditional mutual information $I(X,Y)$. 
\begin{align}
    TE_{X\rightarrow{}Y}=I(Y_t;X_{t-1:t-p}|Y_{t-1:t-p}).    
\end{align}
Where $X_{t-1:t-p}$ indicates the lag 1 to p observations of $X$. Transfer entropy (TE) analysis mitigates some spurious relations by conditioning on the auto-regressive components of the response variable. However, it does not account for spurious relations resulting from surrogacy of third variables. This issue has been explored within the context of TE through the development of causation entropy (CE) \cite{CE}, which introduces additional conditioning on some variable set $\boldsymbol{Z}$, i.e., 
\begin{align}
   CE_{X\rightarrow{}Y}=I(Y_t;X_{t-1:t-p}|Y_{t-1:t-p},\boldsymbol{Z}_{t-1:t-p}). 
\end{align}
The CE equivalent to the full multivariate modelling seen in VAR would be to set this variable set $\boldsymbol{Z}$ to be the entire remaining set of nodes. However, incorporating more information significantly increases model complexity, potentially rendering the estimation process infeasible. As a result, the focus of research continues to be TE, where the lower dimentionality allows for more complex estimators. 

It can be demonstrated that TE simplifies to Granger causality in the case of bivariate vector auto-regression \cite{PhysRevLett.103.238701}. Therefore, when constructing our TE network the assumption of a linear form allows us to apply our previously developed VAR methodology to the two-variable case. In this network, an edge $e_{i,j}$ signifies a statistically significant coefficient $\alpha_1$ in the following bivariate VAR  equation:
\begin{align}
   y_{j,t}=\alpha_0y_{j,t-1}+\alpha_1y_{i,t-1}+c_j+\epsilon_{j,t}  \label{eq:biv_var} . \ 
\end{align}
Reflecting on the three selected network construction methods, we find they all measure linear dependence between our target variables and the lagged realizations of potential covariates. The distinction lies in the degree of conditioning: the correlational graph includes no conditionality; the TE network controls for auto-regressive effects; and the VAR model incorporates full cross-variable conditioning. By comparing the results of these three constructions we hope to gain insight into the practical consequences of the degree of conditioning. 

\subsection{Constructing and analysing networks}

For each of the network construction methods we now have a set of $t_{i,j}$ values that may be arranged into a matrix $T$ such that $T_{i,j}$ is the $t_{i,j}$ value associated with the relationship between lagged values of asset $i$ and current values of asset $j$. For a given critical value $t^*$ we can then replace $T$ with the indicator matrix $T^*= I_{|T|>t^*}$, which can be considered the adjacency matrix $W$ for our estimated network $G$. For the empirical results in section \ref{sec:Results} we use the critical value $t^*=2.57$, which corresponds to a false positive edge detection rate of $\sim1\%$ under the asymptotically normal behaviour.

A common metric of a complex network is the degree $k(i)$ of a given node $i$. For a digraph this comes in two components: the out-degree $k^+(i)=\sum_jW_{i,j}$ specifies the number of outgoing edges, while the in-degree $k^-(i)=\sum_jW_{j,i}$ counts incoming edges. From the set of degrees $k$ in our network we may construct degree distributions $P^+(k)$ \& $P^-(k)$, which describe the probability of a randomly selected node having degree equal to $k$ \cite{Newman2010}.

A common metric for measuring the tendency of the nodes to form cliques is the clustering coefficient. 
We follow the variant used in \cite{clustering}, which defines clustering the clustering $c(i)$ of a node as follows: 
\[c(i)=\frac{(W+W^T)^3_{ii}}{2[k(i)(k(i)-1)-k^{<>}(i)]}, 
\]
\noindent where $k(i)= k^+(i)+k^-(i)$ and $k^{<>}(i)$ is the reciprocal degree, or the number of nodes $j$, such that there is both link $e_{i,j}$ and $e{j,i}$. The clustering of a graph $C(G)$ can be calculated as the mean node clustering $1/N\sum_ic(i)$.

Another common network metric is the eigenvalue centrality (EC) of the nodes, which is a generalised measure of network influence. When considering the influence of a node the eigenvalue centrality considers not only how many outgoing links are present, but also the influence of the target nodes. Nodes receive a larger boost in their EC score when forming links to important nodes. Mathematically, this value is found by solving the eigenvalue problem for the networks adjacency matrix
$A\boldsymbol{x}=\lambda\boldsymbol{x}$.  
The centrality scores are determined by the eigenvector $x$ associated with the largest eigenvalue $\lambda$. 

\section{Analyzing networks}
\label{sec:Results}

\begin{figure*}[h]
     \centering
     \begin{subfigure}[b]{0.34\textwidth}
         \centering
         \includegraphics[width=\textwidth]{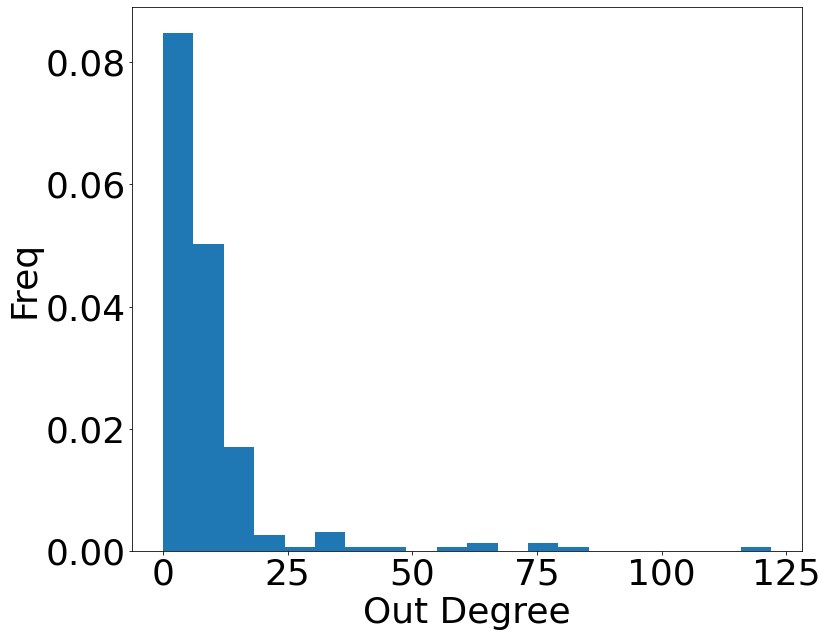}
         \caption{Out-degree distribution histogram. }  
         \label{fig:out_deg_dist}
     \end{subfigure}
     \hfill
     \begin{subfigure}[b]{0.32\textwidth}
         \centering
         \includegraphics[width=\textwidth]{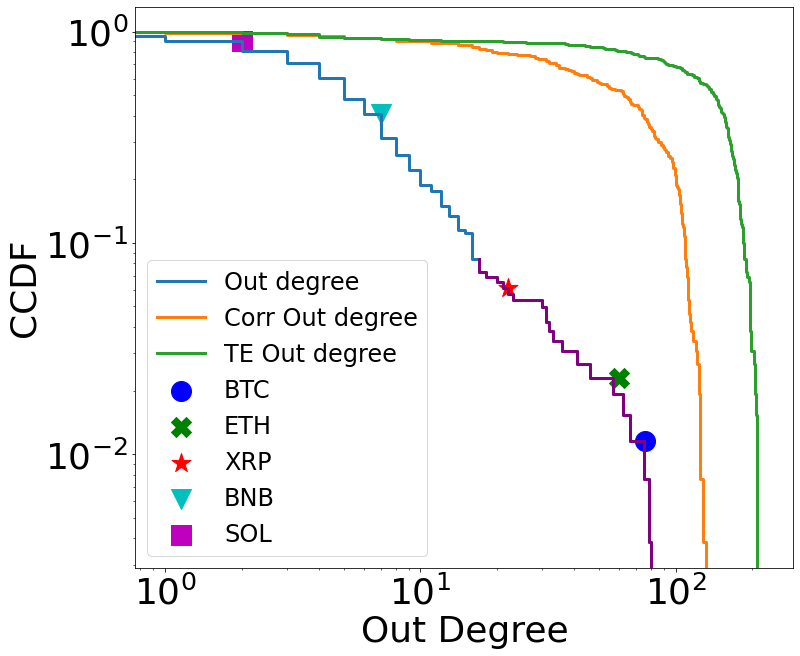}
         \caption{Out-degree CCDF.}
         \label{fig:out_deg_ccdf}
     \end{subfigure}
     \hfill
     \begin{subfigure}[b]{0.32\textwidth}
         \centering
         \includegraphics[width=\textwidth]{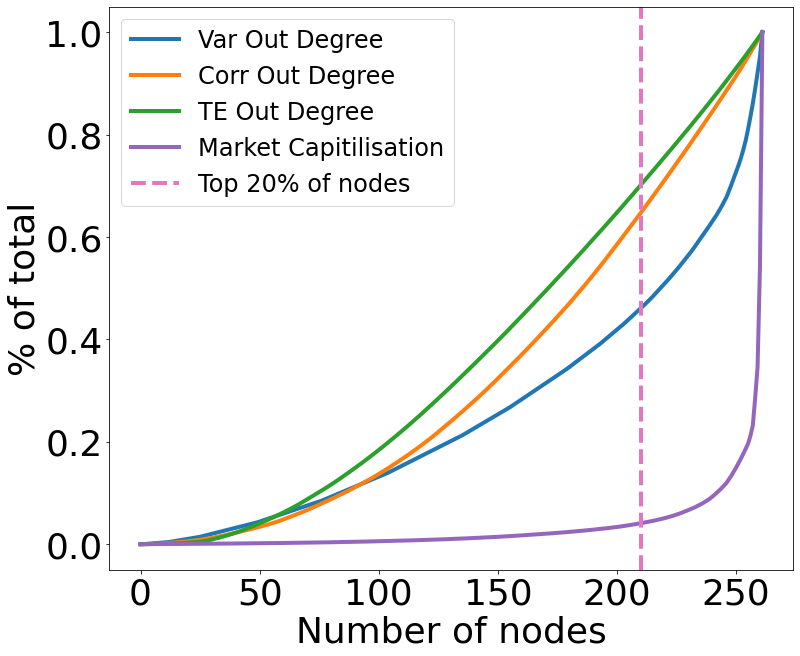}
         \caption{Out-degree Lorenz curve.}
         \label{fig:lorenz_out_deg}
     \end{subfigure}
     \label{fig:out}
     \vspace{-0.25cm}
     \caption{Out-degree distributional figures. Figure \ref{fig:out_deg_dist} displays VAR network information, while \ref{fig:out_deg_ccdf} and \ref{fig:lorenz_out_deg} also displays the alternative networks. Figure \ref{fig:out_deg_ccdf} marks the top 5 capitalised coins, and shows the top 20\% of nodes with a purple line.}
\end{figure*}

\begin{figure*}[h]
     \centering
     \begin{subfigure}[b]{0.34\textwidth}
         \centering
         \includegraphics[width=\textwidth]{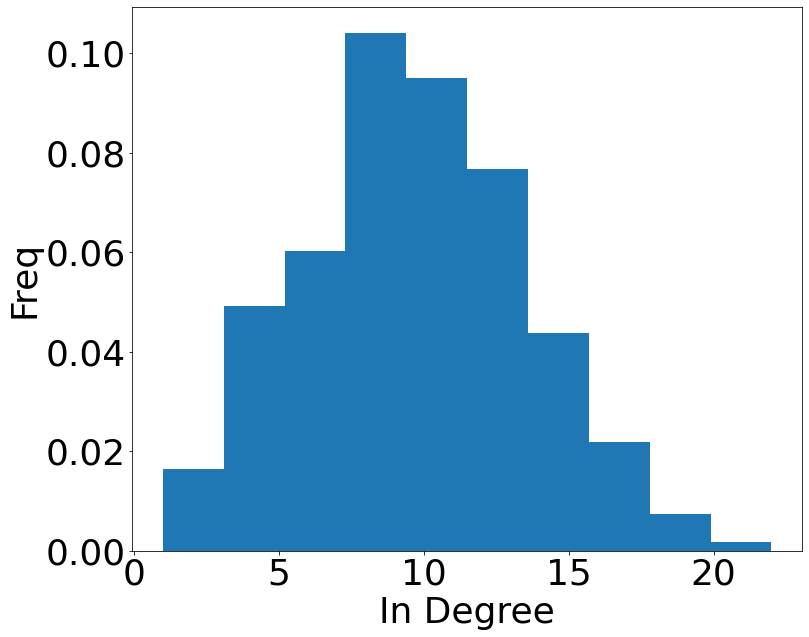}
         \caption{In-degree distribution histogram.}
         \label{fig:in_deg_dist}
     \end{subfigure}
     \hfill
     \begin{subfigure}[b]{0.32\textwidth}
         \centering
         \includegraphics[width=\textwidth]{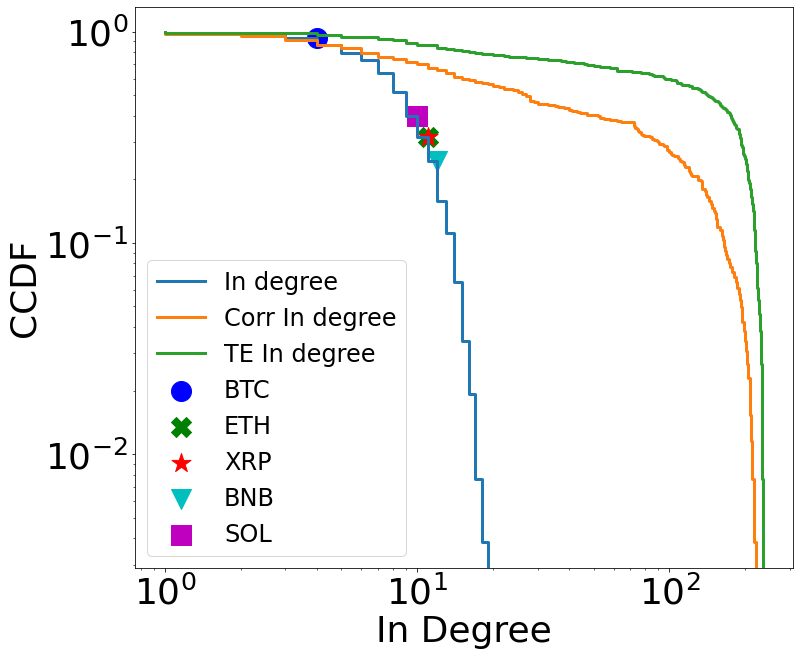}
         \caption{In-degree CCDF.}
         \label{fig:in_deg_ccdf}
     \end{subfigure}
     \hfill
     \begin{subfigure}[b]{0.32\textwidth}
         \centering
         \includegraphics[width=\textwidth]{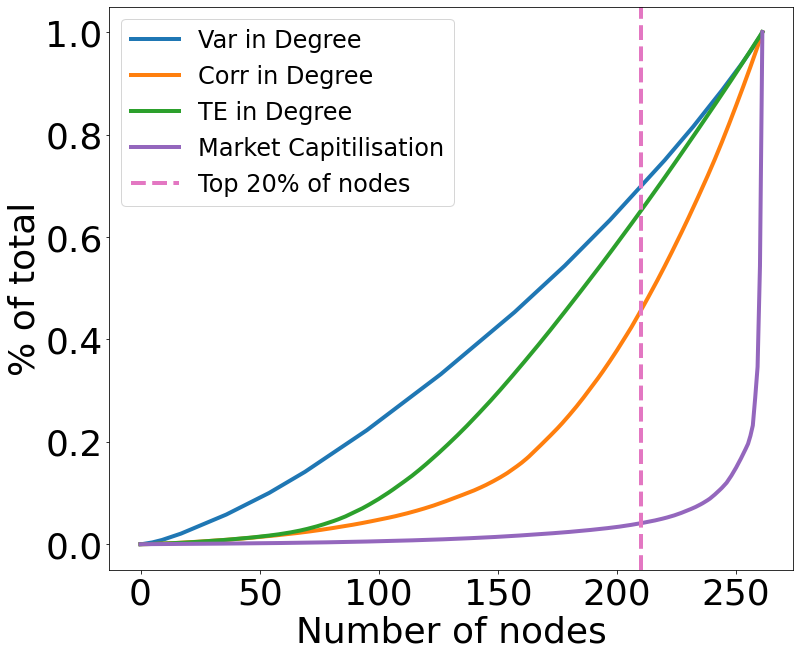}
         \caption{In-degree Lorenz curve.}
         \label{fig:lorenz_in_deg}
     \end{subfigure}
     \label{fig:degree_dist}
     \vspace{-0.25cm}
     \caption{In-degree distributional figures. Figure \ref{fig:in_deg_dist} displays VAR network information, while \ref{fig:in_deg_ccdf} and \ref{fig:lorenz_in_deg} also display the alternative networks. Figure \ref{fig:in_deg_ccdf} includes markers for the top 5 capitalised coins.}
\end{figure*}

The network resulting from our VAR methodology applied to cryptocurrency data is displayed in Figure \ref{fig:network}.
We begin analysing this network by looking into several structural properties such as the node degrees and associated distributions.

\subsection{Comparison of networks}

Node degree histograms for the VAR network are plotted in Figures \ref{fig:out_deg_dist} and \ref{fig:in_deg_dist}, with associated log complementary cumulative density functions (CCDFs) in \ref{fig:out_deg_ccdf} and \ref{fig:in_deg_ccdf}, along with the CCDFs for the alternative network constructions. 
Figure \ref{fig:in_deg_dist} shows that the in-degree distribution $P^-(k)$ is relatively symmetrically distributed around mean 9.79 with relatively low dispersion (SD 3.68), achieving a maximum value of 22. 
The out degree displays substantial right skew in \ref{fig:out_deg_dist}, with mean (9.79) substantially greater than the median (6). 
However, the out degree distribution decays sufficiently fast to avoid being a power law distribution, which would generate an approximately linear log-log CCDF 
in Figure \ref{fig:out_deg_ccdf}. 

We also investigate the relation between market capitalisation and node properties. Table \ref{tbl:varnet} displays basic network statistics, as well as the Spearman rank coefficient and associated $p$ values for several node properties. Statistically significant correlations exist between market capitalisation and out degree, clustering and centrality. In degree is not significantly correlated with capitalisation, however it is correlated with out degree ($\rho = 0.343$, $\textbf{p}=1.29$e$-8$).

Comparatively, the CCDFs of the correlational and TE networks displayed in Figures \ref{fig:in_deg_ccdf} and \ref{fig:out_deg_ccdf} show reduced variation between the in-degree and out-degree distributions. 
Table \ref{tbl:varnet} reveals that the correlation network has a significantly higher number of edges, with the In-degree distribution emerging as the degree type with higher variance. We observe an increase in both absolute effect size and statistical significance of our correlations, with an introduced negative correlation between In-degree and capitalisation. 
The TE network displays comparable characteristics, though with a further increased edge count. The previous correlation between clustering and capitalisation no longer exists, and the negative correlation between in-degree and capitalisation exhibits a substantial rise. 
\begin{table}[ht!]
    \caption{Network Metrics: mean, median, standard deviation $\sigma$ , Spearman's rank correlation ($\rho$) with market capitalisation and associated \textbf{p} value for capitalisation independence.}
    \vspace{-0.25cm}
    \def\arraystretch{1.1}
    \resizebox{\columnwidth}{!}{
        \centering
        \begin{tabular}{l l S[table-format=1.4] S[table-format=1.4] S[table-format=2.4] S[table-format=1.4] S[table-format=1.6]} 
        \toprule
        & \bf{Attribute} & \bf{mean} & \bf{median}  & $\sigma_{attribute}$ & $\rho$ & \bf{p}\\
        \midrule
        \multirowcell{4}{\rotatebox[origin=c]{90}{VAR}} &Out-Deg. & 9.79  & 6 & 13.6 & 0.193 & 0.00175\\
         & In-Deg.  & 9.79 & 10 & 3.68 & 0.0896 & 0.14\\
         & Clust.  & 0.0972 & 0.0855 & 0.0603 & 0.129 & 0.0368\\
         & Central. & 0.324 & 0.326 & 0.0964 & 0.207 & 0.000753 \\
         \hline
        \multirowcell{4}{\rotatebox[origin=c]{90}{Correlation}} & Out-Deg. & 62.1  & 64 & 36.8 & 0.386 & \num{1.12e-10} \\ 
        & In-Deg.  & 62.1 & 28 & 64.03 & -0.170 & 0.00592\\
        & Clust.  & 0.361 & 0.379 & 0.0732 & 0.131 & 0.0344\\
        & Central. & 0.528 & 0.545 & 0.00832 & 0.381 & \num{1.83e-10} \\ 
        \hline
        \multirowcell{4}{\rotatebox[origin=c]{90}{TE}} & Out-Deg. & 124  & 143 & 58.7 & 0.375 & \num{4.1e-10} \\ 
        & In-Deg.  & 124 & 143 & 82.3 & -0.383 & \num{1.61e-10} \\ 
        & Clust.  & 0.598 & 0.610 & 0.0610 & 0.0127 & 0.834\\
        & Central. & 0.656 & 0.684 & 0.117 & 0.373 & \num{4.8e-10} \\ 
        \bottomrule
        \end{tabular}
        \label{tbl:varnet}
        }

\end{table}

Comparing figure \ref{fig:out_deg_ccdf} to \ref{fig:market_cap_ccdf} we see that the VAR out-degree is significantly less power-tailed than market capitalisation, with a faster decaying survival function and ultimately less emphasis on extreme outliers. Another way of considering a power-tail distribution is that a single observation, or small subsets may contribute very large proportions of the mean value of the random variable (or equivalently, of the sum of all values). One way to visualise this relative dominance of specific nodes along an attribute is to plot the Lorenz curve, showing the cumulative sums when ordering nodes based on these attributes. These curves allow us to visualise how much of a total, such as total market capitalisation of the market is contained within the top say, 20\% of the network. From these Lorenz curves one can derive the Gini coefficient, a metric that numerically describes the amount of curvature, hence concentration, of the relevant value. 

Figures \ref{fig:lorenz_out_deg} and \ref{fig:lorenz_in_deg} show Lorenz curves for the market capitalisation, out degree and in degree for all networks, alongside a vertical line at the 52 highest nodes (Top ~20\% of the network). 
An extreme Pareto effect exists for market capitalisation, with the top 20\% of nodes having ~96\% of the total market capitalisation (Gini = 0.96). 
Our measure of influence, node degree, is much more evenly dispersed (Gini = 0.48) with the top 20\% containing around 56\% of outgoing links; in-degree is near-linearly dispersed (see the histograms in Figure \ref{fig:in_deg_dist}. 
While influence in our network inherits some degree of heavy scaling, it is nowhere near as `winner takes all' as the capitalisation figures. 
The in degree curves in Figure \ref{fig:lorenz_in_deg} show that both alternative networks have increased curvature (Gini = 0.21 vs 0.56 \& 0.38), indicating more uneven allocation of in degrees, and matching the increased variance of these distributions in Table \ref{tbl:varnet}. 
The out degrees in \ref{fig:lorenz_out_deg} show the opposite effect, with reduced curvature (Gini = 0.48 vs 0.34 \& 0.26). 
   
Surrogate variable effects potentially explain these qualitative changes. 
For example, assume that a coin $j$ is highly dependent on another coin $k$ in the full multivariate model. 
For the non-multivariate networks we may expect to see a potential link $e_{i,j}$ between $j$ and all variables $i$ that are highly correlated with $k$. 
This would lead to many new links, starting at different variables and pointing towards the same target $j$. 
Such effects would significantly increase $j$'s in degree, and slightly increase many variables' out degree, leading to the type of Lorenz curve changes we observe. 

\begin{figure*}[h!]
     \centering
     \begin{subfigure}[h!]{0.48\textwidth}
         \centering
         \includegraphics[width=0.85\textwidth]{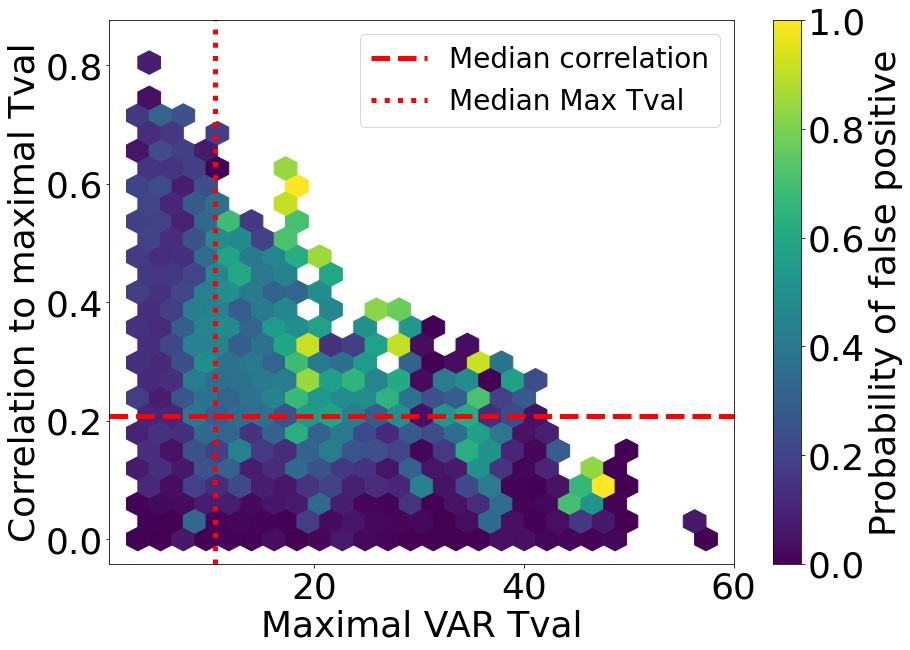}
         \caption{Correlation Network.}
         \label{fig:Corr_triangle}
     \end{subfigure}
     \hfill
     \begin{subfigure}[h!]{0.48\textwidth}
         \centering
         \includegraphics[width=0.85\textwidth]{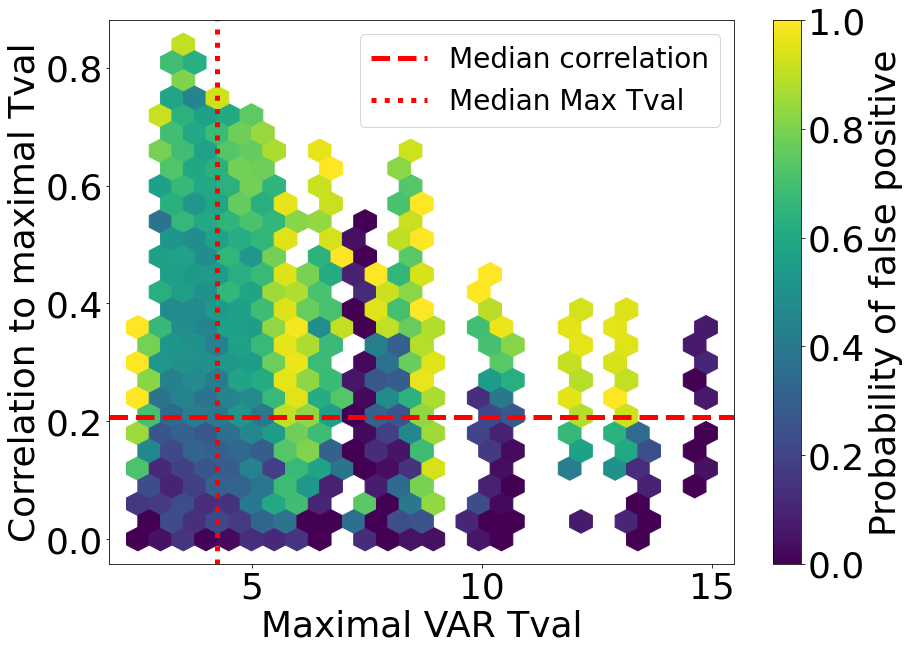}
         \caption{TE Network.}
         \label{fig:TE_triangle}
     \end{subfigure}
     \vspace{-0.25cm}
     \caption{The `False Positive' rate against large correlated T values. Red lines indicate median value along the specified axis, with coloration indicating the probability of a false positive link, treating the VAR network as ground truth.}
\end{figure*}

To investigate whether introduced edges are related to simultaneous correlation effects we visualise the relationship between significant causal edges, simultaneous correlations and the 'false positive' (FP) rate, where we take the VAR network as ground truth. 
For each given target variable $j$, we find the set of t values associated with incoming edges $\{t_{i,j} \: \forall \ i \in V\} = T_{,j}$. We then determine the source node $k$ corresponding to the maximal element of this set $t_{k,j}=\max(T_{j,})$, and simultaneous correlation $R(0)_{i,k}$ for each source $i\neq k$ in $T_{j,}$. We include the additional constraint $k\neq j$ for the TE network, as corresponding $t_{i,j}$ values are already conditioned on autoregressive effects in $j$. Figures \ref{fig:Corr_triangle} and \ref{fig:TE_triangle} plot $R(0)_{i,k}$ against $t_{k,j}$, with cell coloration based on the FP probability (plotting only cells with $>10$ observations).

When a target $j$ has significant causal dependence on a variable $k$ (high $t_{k,j}$), potential sources $i$ with high simultaneous correlation $R(0)_{i,k}$ have significantly elevated FP probability. Considering the upper right region, with lines determined by the median $R(0)_{i,k}$ and median $t_{k,j}$, there is a greater FP propensity across both network types. While this variable effect occurs in both figures,  FP probabilities are lower in the correlation network, as implied by the overall lower number of edges (Table \ref{tbl:varnet}). 
Assuming the VAR network is the true label, Table \ref{table:classification_metrics} shows several classification/co-occurrence metrics for these alternative constructions. 
While the TE network shows higher recall, this is  offset by reduced precision values, such that the combined F1 score is higher for the correlation network. 
Both networks are poor classifiers with respect to the VAR network, hence, the correlation network, containing fewer links, has overall better performance. 

\begin{table}[h!]
    \centering
    \caption{Classification metrics of alternative network constructions, treating the VAR network as ground truth.}
    \vspace{-0.25cm}\label{table:classification_metrics}
    \def\arraystretch{1.1}
    \resizebox{\columnwidth}{!}{
    \begin{tabular}{l c c c c c c c} 
    \hline
    \bf{Network} & \bf{TP} & \bf{TN} & \bf{FP}  & \bf{FN} & \bf{Precision} & \bf{Recall} & \bf{F1} \\
    \hline
    \bf{Corr Net} & 0.012 & 0.74 & 0.23 & 0.026 & 0.051 &  0.32 & 0.0859 \\
    \bf{TE Net} & 0.020 & 0.51 & 0.46 & 0.018 & 0.041 & 0.52 & 0.0773\\
    \hline 
\end{tabular}%
}
\end{table}

\subsection{Reductive Experiment}

\begin{figure*}
    \centering
    \includegraphics[width=\textwidth]{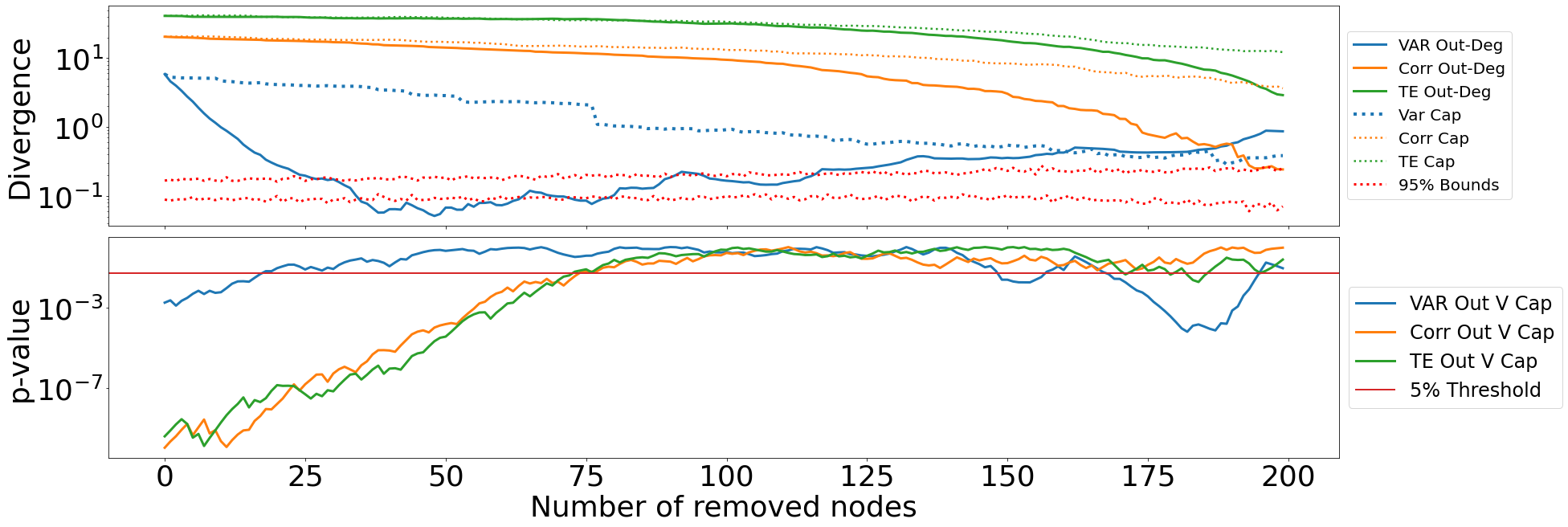}
    \vspace{-0.25cm}
    \caption{Out-degree KL divergence and market cap correlation vs removed nodes. Red dotted lines mark the 95\% confidence interval for binomial comparison, with the solid red line showing the 5\% statistical significance threshold. The upper image's dotted lines show divergence results with removal ordering based on capitalization.}
    \label{fig:reduc_exp}
\end{figure*}

There are significant outliers in both out-degree and market capitalisation, and correlations between these attributes and several network activity metrics. We may ask whether there exists a VAR subnetwork that contains a large amount of the meaningful activity. 
Formally, We define a hypothesis test that there exists some subset of the graph $G^*$ that contains most of the relevant ``information", such that the graph without this subset is mostly structure-less, and has minimal relevant correlations. To test this hypothesis, we conduct a reductive experiment by sequentially removing the most influential nodes (highest out degree). 
This aims to determine how rapidly the remaining network becomes ``low information'': out degree distributions resembling pure random allocation of edges and  statistically insignificant correlations.

During each round $m$ of this process, we compare the reduced graph $G_m$ against a theoretically structureless/null graph by caclulating the Kullback–Leibler (KL) divergence $D_{KL}(P||Q)$ of the out-degree distribution. 
KL divergence measures the difference between two probability distributions, in this case, the empirical out-degree distribution of the reduced graph $\hat{P}^{+}_m(k)$, and the out-degree distribution $Q_m$ of an Erdos-Renyi random graph (fitted with $\hat{p}_m=|E_m|/(N-m)^2$).
We track changes in the signficance of network correlations by re-calculating and documenting the Spearman $p$ values at each step of the process. 
Crucially, we emphasize that our primary focus is on how quickly the divergences approach 0, as they are anticipated to diverge again beyond this point. 
If we remove high value nodes from something that resembles a binomial network, it will by definition stop being a binomial network.

Figure \ref{fig:reduc_exp} shows that for the VAR network the majority of out degree distributional deviation from the random graph has disappeared after removing the $\sim35$ most influential nodes. 
Similarly, the lower panel shows that capitalisation correlation is generally below the significance threshold after removing $\sim30$ nodes. When nodes are removed based on capitalisation, the decay is significantly slower.
Comparatively, both alternative network constructions show much slower information decay, indicating that these networks do not contain an equivalently concentrated subset of network activity.

This shows that the majority of the network's information is concentrated within a relatively small subset, with approximately 20\% of nodes containing the majority of correlations and deviations from random structures. This subset cannot be simply characterised as the 20\% highest capitalised coins, indicating the the dynamics of influence is only weakly explained by capitalisation, and their exists an independent, latent subset of influencer coins. The observed concentration of information in this subset underscores the importance of these high influence nodes in determining the overall behavior and stability of the cryptocurrency market.

\section{Conclusion}
\label{sec:Conclusion}
This study demonstrates the effectiveness of multivariate linear models in constructing informative and economically intuitive cryptocurrency networks. 
We considered several network constructions representing different levels of cross conditioning on asset histories, and found that the non-multivariate frameworks display substantial surrogate variable effects. 
This implies that while models can be of a simple statistical nature, there is a computational burden associated with multivariate effects that may hinder complex models' applicability due to the rapidly growing number of parameters. 
In a Pareto-like manner, only a small portion of the network accounts for the majority of the overall structure. 
Our multivariate model reveals significant correlation between both forms of node degree, clustering and market capitalisation, confirming the relation between capitalisation and influence in the dynamics of the cryptocurrency market. 
Further, while market capitalisation plays a crucial role in determining the cryptocurrency network, there exists a subset of nodes with influence surpassing what market capitalization would suggest.
This has implications for market participants, as it highlights the potential presence of hidden influencers that could significantly impact the overall stability and dynamics of the cryptocurrency ecosystem.
This study serves as a baseline for further research into conditioning methodology to control cross-variable effects in causal networks. 
The demonstration of linear causality in terms of expected returns can also be extended into variance and risk modelling frameworks.


\bibliographystyle{ACM-Reference-Format}
\bibliography{my_conference_paper}


\end{document}